# SPECTRAL ELEMENT METHODS FOR LIQUID METAL REACTORS APPLICATIONS[1]

**Elia Merzari, Aleksandr Obabko, and Paul Fischer**

Argonne National Laboratory, United States of America

## 1. Introduction

Funded by the U.S. Department of Energy, the Nuclear Energy Advanced Modeling and Simulation (NEAMS) program aims to develop an integrated multiphysics simulation capability "pellet to plant" for the design and analysis of future generations of nuclear power plants. NEAMS embraces a multiresolution hierarchy designing the code suite structure to ultimately span the full range of length and time scales present in relevant reactor design and safety analyses.

Advanced reactors, such as liquid metal reactors, rely on innovative component designs to meet cost and safety targets. Heuristic assumptions typically employed in the safety analysis of the current fleet are not likely be sufficient. In order to span a wider design range, advanced modeling and simulation capabilities that rely on minimal assumptions (i.e., fewer closure models) plays an important role in optimizing the design.

Over the past several years the NEAMS program has developed the integrated multiphysics code suite SHARP [1] aimed at streamlining the prototyping of such components. The goal of developing such a tool is to perform multiphysics neutronics, thermal/fluid, and structural mechanics modeling of various reactor components at user-specified fidelity, with a focus on first principles approaches ("high fidelity").

The thermal-hydraulic philosophy in NEAMS relies on a hierarchical multiscale approach (Fig. 1) consistent with other programs, such as THINS. SHARP focus on the high-fidelity end, aiming primarily at turbulence-resolving techniques such large eddy simulation (LES) and direct numerical simulation (DNS). The computational fluid dynamics code (CFD) selected for SHARP is Nek5000 [2], a state-of-the-art highly scalable tool employing the spectral element method (SEM). In the following we review the method and its implementation in Nek5000 (Section 2). We also examine several applications (Section 3). In Section 4 we summarize our work.

---

[1] *Submitted for publication to the Von Karman Institute of Fluid Dynamics*

We note that Nek5000 is also regularly employed for intermediate-fidelity approaches such as Reynolds-averaged Navier-Stokes (RANS) and for reduced-order models employing momentum sources or porous media, especially when coupled to neutronics modeling.

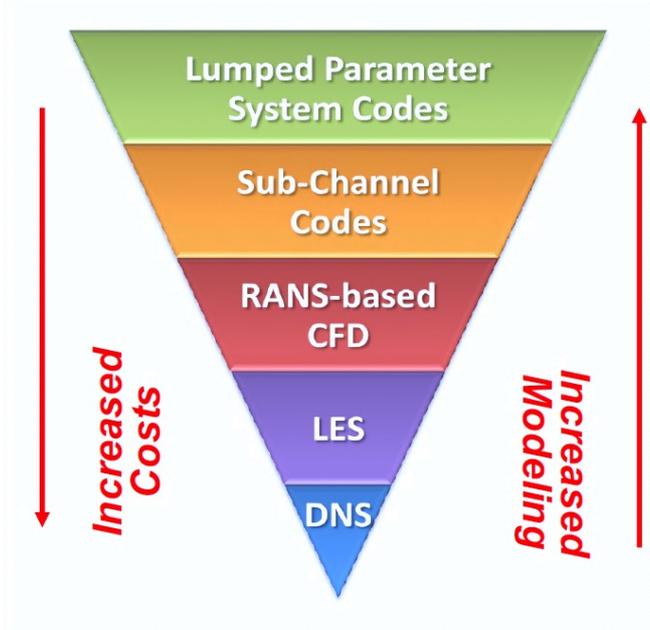

Figure 1: Thermal-hydraulic hierarchy in NEAMS. Nek5000 covers LES, DNS, and RANS.

## 2. Nek5000 and the Spectral Element Method

Nek5000 [2] solves the unsteady incompressible two-dimensional, axisymmetric, or three-dimensional Stokes or Navier-Stokes equations with forced or natural convection heat transfer in both stationary (fixed) and time-dependent geometry. It also solves the compressible Navier-Stokes in the low Mach number regime and the magnetohydrodynamic equations. Additionally, Nek5000 can handle conjugate heat transfer problems. The solution variables for a typical thermal-hydraulic problem are the fluid velocity **u** = ($u_x$, $u_y$, $u_z$), the pressure $p$, and the temperature $T$. Coupling to neutronics and structural mechanics is provided as part of the SHARP suite. The capability to solve multiphase flows and fully compressible flows has been added recently. The equations of interest for the work presented here are primarily the constant property Navier-Stokes equations:

$$\rho \left( \frac{\partial u}{\partial t} + u \nabla u \right) = -\nabla p + \mu \nabla^2 u \qquad (1)$$

$$\rho c_p \left(\frac{\partial T}{\partial t} + u\nabla T\right) = \lambda \nabla^2 T \tag{2}$$

$$\nabla u = 0, \tag{3}$$

where $\mu$ is the molecular viscosity and $\lambda$ is the conductivity. We note that while for most cases presented here both variables are constant, the code can handle spatially and temporally varying properties. The spatial discretization in Nek5000 is based on the spectral element method [3], a subclass of Galerkin methods, or weighted residual methods. The functional space of **u**, *p*, and *T* is described as a polynomial expansion. The foundational idea is to minimize the error of the numerical computation in the energy norm over a chosen space of polynomials:

$$u = \sum_i u_i h_i(x), \tag{4}$$

where *h* represents a Lagrangian polynomial constructed on Gauss-Lobatto-Legendre (GLL) collocation points within the canonical subdomain, $[-1,1]^3$, which is mapped to curvilinear hexahedral (i.e., brick) elements. Stability restricts our attention to certain acceptable classes of basis functions, *h*, in the reference domain. If modal, they are typically linear combinations of orthogonal polynomials. If nodal, the nodes are typically chosen to be the zeros of orthogonal polynomials or, equivalently, quadrature points associated with a particular Gauss quadrature rule—such distributions tend to cluster points toward the ends of the interval in the unit box. Virtually all such combinations ensure stability, that is, reasonably small condition numbers of the resultant matrices. (Examples of unstable bases include monomials and Lagrange interpolants on uniform point distributions.)

Additionally, in the SEM, one typically (though not always) employs quadrature on the underlying GLL nodal points to evaluate integrals, a process that greatly simplifies many of the operator evaluations, resulting in reduced costs and, significantly, a diagonal mass matrix. The GLL quadrature is exact for any polynomial integrand of order up to 2N – 1, and the error incurred is on a par with the truncation error. This property, combined with a tensor product formulation of the differential operator, allows significant reduction in the computational cost per timestep of SEM at high order, which scales linearly with the number of grid points and has only a weak dependence on the polynomial order. On the other hand, the computational cost of standard p-type finite elements scales with the sixth power of the polynomial order.

In Nek5000, the domain is decomposed globally into smaller domains (elements), which are assumed to be curvilinear hexahedra that conform to the domain boundaries. Locally, functions within each element are expanded according to (4), and operators are cast in tensor product form. Temporal discretization is based on a high-order splitting that is third-order accurate in time and reduces the coupled velocity-pressure Stokes problem to four independent solves per timestep: one for each velocity component and one for the pressure. The velocity problems are diagonally dominant and thus easily solved by using Jacobi preconditioned conjugate gradient iteration.

The pressure substep requires a Poisson solver at each step, which is effected through multigrid-preconditioned GMRES iteration coupled with temporal projection to find an optimal initial guess. Particularly important components of Nek5000 are its scalable coarse-grid solvers that are central to parallel performance. For large eddy simulation, the subgrid-scale model employed relies on explicit filtering: the energy content of the highest few wavenumbers is filtered to effect grid-scale dissipation.

Nek5000 employs a pure MPI parallel implementation; and thanks to key choices described above it, is highly scalable. Figure 2 reports a recent study on a Blue Gene/Q supercomputer showing good scaling up to 1,000,000 MPI ranks for a 217-pin wire-wrapped rod bundle.

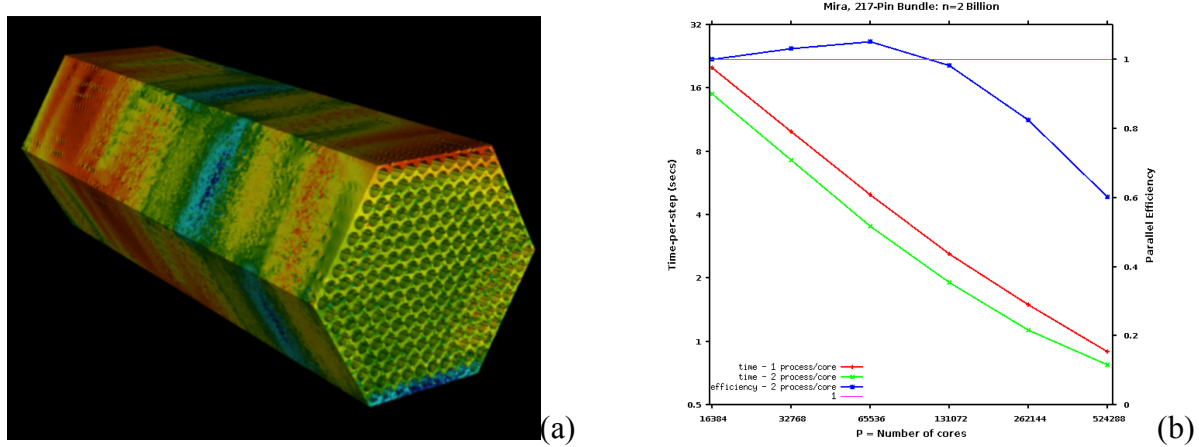

Figure 2: Scaling study in a wire-wrapped rod bundle in which Nek5000 exhibits good strong scaling even at 1,000,000 MPI ranks: (a) pressure distribution in a 217-pin wire-wrapped rod bundle; (b) strong scaling.

The availability of a high-order, flexible, scalable methodology is paramount to perform high-fidelity (DNS/LES) simulations in large-scale complex geometries. While high-order solvers can reduce the requisite number of degrees of freedoms, scalability ensures that the cost per grid point remains low as the problem increases in size.

## 3. Liquid Metal Reactor Applications

In the following we examine several examples of the use of Nek5000 in liquid metal reactor applications.

### 3.1. Fuel Assemblies

We first consider three rod bundle configurations.

3.1.1 Bare Bundles

The axial flow in rod bundles has been the object of several investigations because it is relevant for a variety of industrial applications (e.g., heat exchangers, nuclear reactor cores). For tight configurations (pitch-to-diameter ratio smaller than 1.1) the large difference in velocity within the cross section creates the possibility of a Kelvin-Helmholtz instability and the generation of a vortex street in the gap ("gap instability"). Considerable work with Nek5000 was done aiming at improving our understanding of the gap instability in small and large bundles [4, 5]. In Fig. 3 we present a large-scale wall-resolved LES simulation of a 37-pin hexagonal lattice rod bundle at a pitch-to-diameter ratio of 1.12, typical of current advanced reactor designs. Although this configuration does not typically generate a vortex street in the center of the domain, the vortices can be observed in the proximity of the canister wall enclosing the rod bundle if the edge subchannels are sufficiently small.

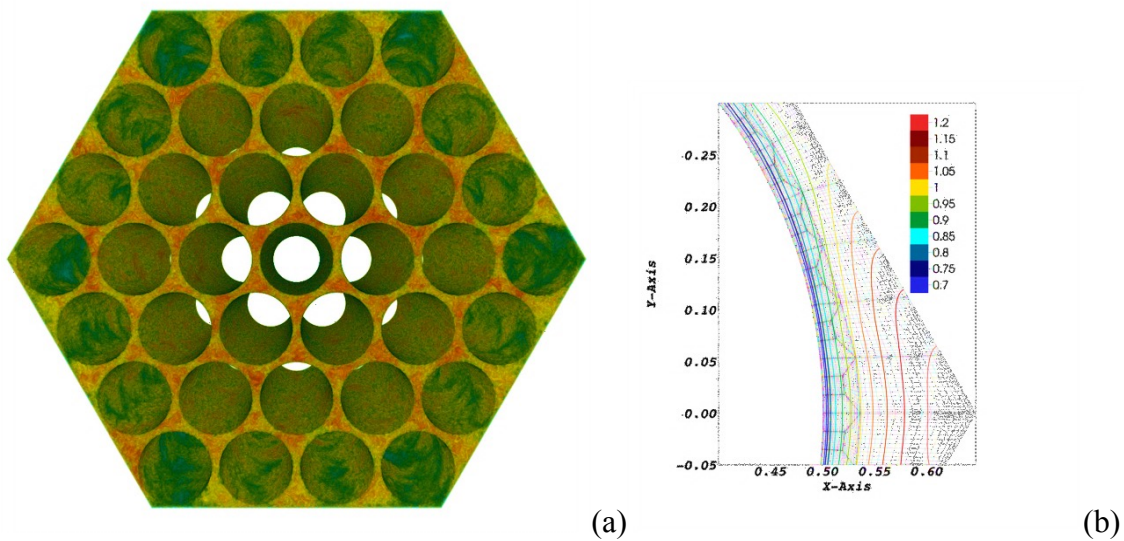

(a)                                                                    (b)

Figure 3: Nek5000 simulation in a 37-pin bare rod bundle: (a) typical mesh for a portion of single pin and averaged velocity distribution; (b) velocity magnitude for wall-resolved LES.

Figure 3a depicts the elements and the actual collocation points for the coarse mesh. The discretization is designed to allow for at least one point near the wall at y+<1 and five points within y+<10. Periodic boundary conditions are applied in the streamwise direction. The length in the streamwise direction is set to 10 D for the 37-pin rod bundle case. This distance is deemed sufficient to describe the coherent structures, while limiting the computational cost. The importance of the streamwise length in these simulations has been discussed in several publications. No-slip boundary conditions are applied at the wall. The azimuthal mesh requirements are finer in rod bundles when compared with channel flow because of the need to capture the azimuthal variation of the wall shear stress. This issue has been discussed at length in [6], demonstrating that using higher-order polynomials as well as a higher resolution near the pin walls leads to significantly better results.

In the LES of rod bundles, the cost of reproducing accurately the wall dynamics and wall shear stress is high and requires optimization. In fact it constitutes the primary cost in wall-resolving LES and supersedes the cost of geometric complexity, as discussed below. For the 37-pin bundle case, run at a Reynolds number of approximately 60,000, over 8 billion collocation points (at 19th polynomial order) were judged necessary in order to achieve accuracy for the wall shear stress. More than 500,000 MPI ranks were used to carry out the simulations (costing approximately 60,000,000 CPU-hours). The machine used was the IBM Blue Gene/Q system Mira at Argonne National Laboratory. The parallel efficiency for this problem proved excellent, exceeding 75% at 500,000 MPI ranks. The simulation was run for over 10 turnover times (TOTs) before collecting turbulence statistics (which included higher-order momentums such as skewness and kurtosis) for 10 additional TOTs.

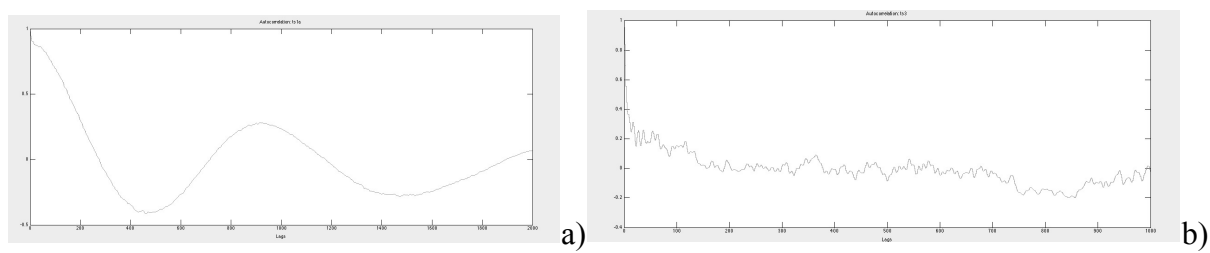

Figure 4. Time series analysis for 37-pin rod bundle: (a) autocorrelation of cross velocity in a point near the canister wall; (b) autocorrelation of cross velocity for point in the middle of the domain.

The simulation shows that for sufficiently small distance between outer pins and canister walls, a strong difference exists between the dynamic behavior in the interior of the channel and the exterior. This difference has been discussed extensively in [4] and indicates the complexity of the flow in rod bundles.

It is highlighted in Fig. 4, where the autocorrelation of the cross-velocity signal is compared in different parts of the domain. Near the canister wall the autocorrelation exhibits the behavior typical of a gap vortex street, while in the center of the domain it presents the typical profile of incoherent turbulence (low correlation). The averaged velocity is shown in Fig. 5a.

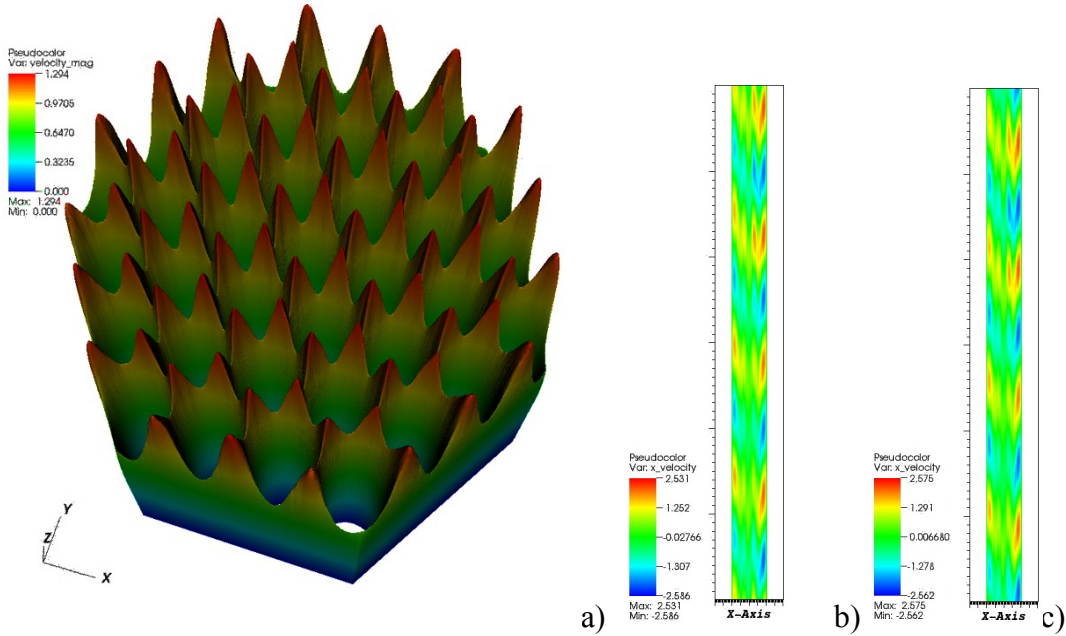

Figure 5. Averaged velocity data for the 37-pin rod bundle case: (a) velocity magnitude 3D plot, typical POD mode pair, for 37-pin rod bundle case with regional clipping, velocity in the x direction and cross section at y=3.42 D; (b) mode 3; (c) mode 4.

One of the key methods used to identify the gap vortex street is proper orthogonal decomposition (POD) [7]. It allows for the recognition of the most energetic modes of turbulence by performing an eigenvalue decomposition of the covariance matrix. A set of M snapshots of the flow field is collected, and the following eigenvalue-eigenvector problem is solved to obtain the POD eigenmodes:

$$C_{mn} a_{n,i} = \lambda_i a_{n,i} , \qquad (5)$$

where $a_{n,i}$ are the real coefficients that define the eigenmodes and $\lambda_i$ are eigenvalues; no summation is performed over $i$. The matrix C (size MxM) is defined as

$$C_{mn} = \frac{1}{M}\left(\vec{u_m}, \vec{u_n}\right), \tag{6}$$

where $\vec{u}_m$ are snapshots of the velocity field. Each eigenmode can be then be computed by using

$$\vec{\sigma_i}(\vec{x}) \cong \sum_n a_{n,i} \vec{u_n}(\vec{x}). \tag{7}$$

In order to examine the local structures, a regional clipping operator L can be applied. The focus can be any region in the domain. Equation (6) becomes

$$C_{mn} = \frac{1}{M}\left(L\vec{u_m}, L\vec{u_n}\right), \tag{8}$$

where *L* will generally be zero, the velocity vector where clipping is being applied.
Over 3,000 snapshots have been collected, and local clipping has been applied to them. This approach has allowed localizing the coherent structure in the outer subchannels. Eigenvectors in this region exist in pairs (Fig. 5b and Fig. 5c), a characteristic of traveling waves [4]. POD applied without clipping produces a less clear structure and is of limited use in identifying and studying the gap vortex street.

3.1.2 Wire-Wrapped Rod Bundles

Most liquid metal-cooled fast reactor designs employ wire wraps as spacers between the individual pins in a rod bundle. Although many experiments have been performed in the past, Roelofs et al. [8] clearly demonstrate that CFD-grade validation data is still not widely available. New thermal-hydraulic experiments have been undertaken to fill this gap, as discussed later in this section. In the past, collaborations were established to compare results from RANS approaches with data from high-fidelity LES in a blind benchmark [9]. In fact, both DNS and LES can provide high-fidelity reference data for comparison with more pragmatic RANS or hybrid approaches.
Such comparison efforts led to the conclusion that a substantial agreement between commonly used models (such the SST k-ω model) and LES can be reached for integral quantities. An example is shown in Fig. 6 with a comparison for the cross flows as a function of the axial coordinate at an arbitrary subchannel location in a 7-pin wire-wrapped rod bundle. The agreement is within a few percentage points and is considered acceptable. Similar agreement was observed in other subchannels.

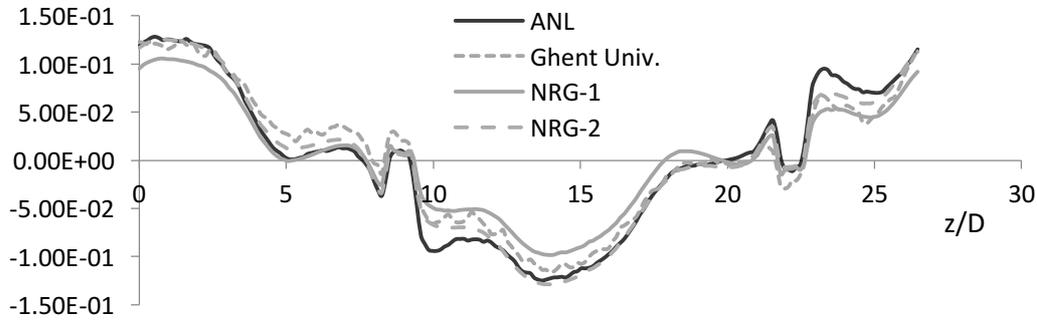

Figure 6. Comparisons of the cross flows in arbitrary subchannel between different codes and models. The ANL data represents an LES performed with Nek5000; the Ghent University and NRG-2 submissions used SST k-ω with respectively FLUENT and STAR-CCM+; and the NRG-1 submission used a cubic k-ε model.

While code-to-code comparisons such as the one described in [9] and shown in Fig. 6 are essential for continued code development, validation through experiments remains necessary. As part of a recent collaboration with industry and academia, Argonne National Laboratory conducted extensive simulations for a 61-pin wire-wrapped rod bundle aimed at CFD validation. The simulations were then compared with results from a particle image velocimetry (PIV) experiment conducted at Texas A&M [10].

To accelerate convergence and optimize the use of limited computational resources, we began with low resolution (i.e., low polynomial order) and Reynolds number of $N$=3 and Re=100, respectively, and scaled the runs up to $N$=8 and Re=40,000. The Reynolds number in the present discussion is based on pin diameter, D, and is equivalent to the experimental conditions. We note that in the case of the experimental bundle, $D/D_h$=2.055, where $D_h$ is the hydraulic diameter. Figure 7 shows the computational mesh for the Texas A&M cases with $N$=3 with a detail highlighting the flattening of the wire. Production runs have been conducted for the target Reynolds number at $N$=7. Figure 8 shows the LES solution at Re=10,000 as flow undergoes transition to fully turbulent flow.

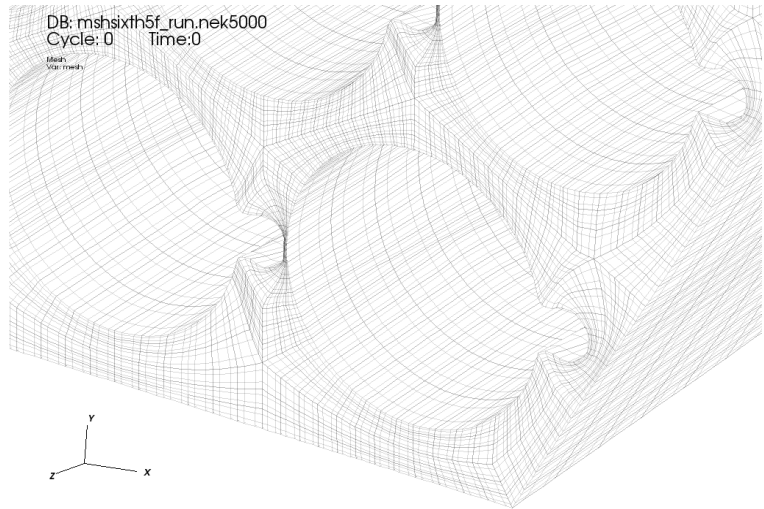

Figure 7: Nek5000 LES mesh for a wire-wrapped rod bundle.

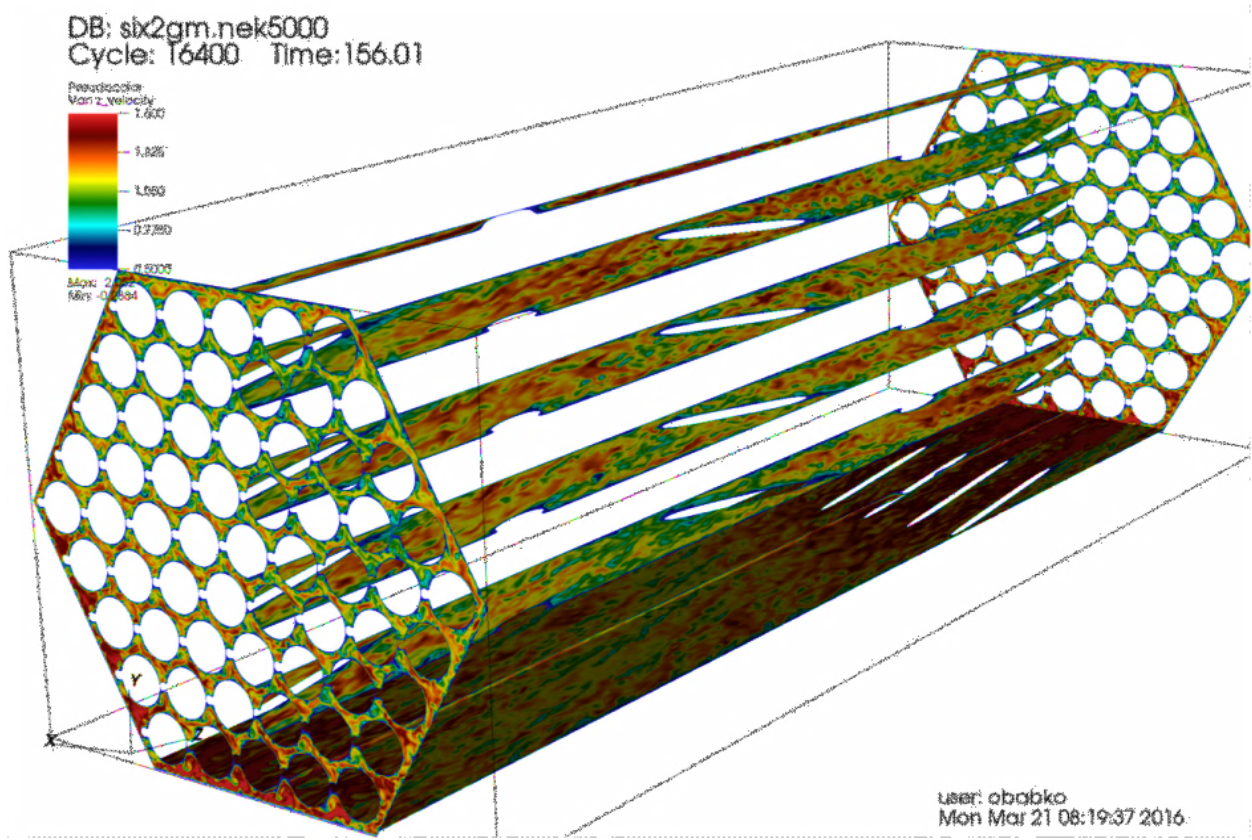

Figure 8: Nek5000 LES instantaneous axial velocity at Re=10,000.

A preliminary comparison in one of the central channels is shown in Figure 9. The agreement is acceptable but will require additional work for a comprehensive evaluation. Uncertainty due to laser position and thickness are not taken into account. Additional tests in smaller bundles are under way and will help further qualify Nek5000 for the simulation of the flow in wire-wrapped rod bundles. We note that an LES calculation of the full height with conjugate heat transfer was performed. This calculation, with over 12 billion grid points, is one of the largest ever performed with Nek5000.

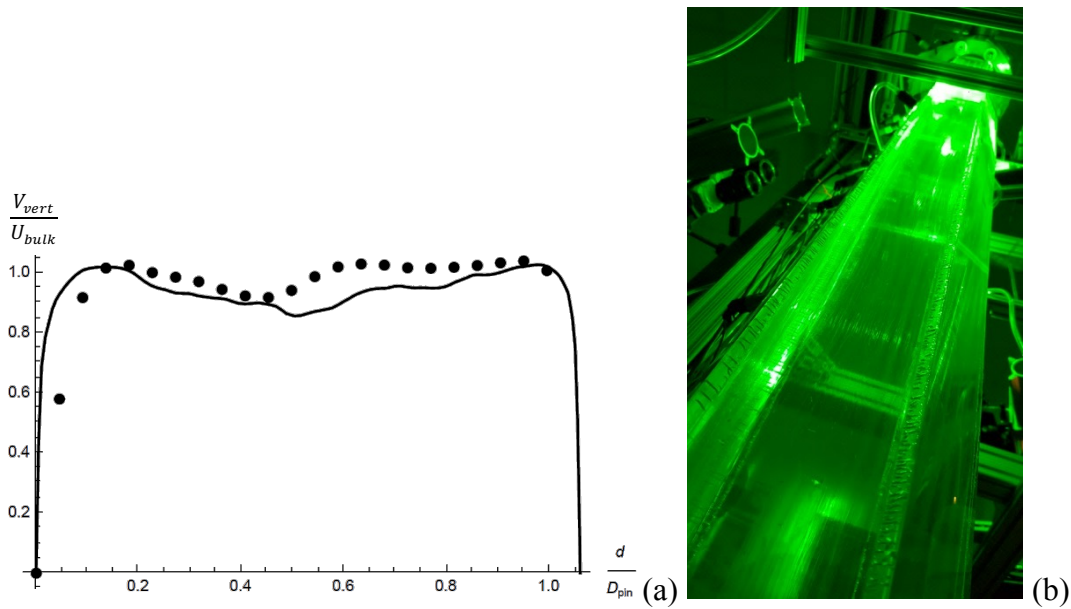

Figure 9: (a) Nek5000 LES (solid line) comparison with experimental data (dots) for a 6-pin rod bundle in a central channel (streamwise velocity); (b) image of the experimental bundle.

3.1.3 Other Rod Bundles

Grid spacers are important components of nuclear reactor fuel assemblies because they allow separating the rods while enhancing turbulence. In liquid metal reactors the typical spacing devices are wire wraps as discussed in Section 3.1.2. Often, however, other spacing devices are considered. A series of LES was conducted for a 19-pin SFR rod bundle with simplified spacers demonstrating that these devices not only are not ideal from the point of view of mixing but also present a considerably higher pressure drop when compared with wire wraps [11].

Moreover, if tight lattice rod bundles are considered, such spacing devices present the same large-scale structures described in Section 3.1.1. Results of the 19-pin rod bundle calculation are discussed in detail in [11].The discretization is designed to allow for at least one point near the wall at y+<1 and five points within y+<10. The simulations were carried out with up to 250,000,000 collocation points (40,000 elements/pin), which were found through a mesh refinement study to be sufficient to describe the flow field accurately.

Figure 10 shows the most energetic POD mode obtained by processing approximately 1,000 snapshots. Regional clipping has been applied, and the focus is the region around the central pin. The mode presents a strong similarity with the principal mode observed in previous analysis (Section 3.1.1). It appears to be less regular, however, because of a combination of the effect of regional clipping and the additional complexity brought by the gap vortex network. Additional POD, focused on larger or different regions of the domain, will need to be performed in order to examine in more detail the dynamics of the vortex network. An additional conclusion is that although spacers introduce complexity into the gap vortex network dynamics, they do not preempt its development. In fact, honeycomb spacers cannot be used to prevent the gap vortex street from developing, as shown in a recent study [11].

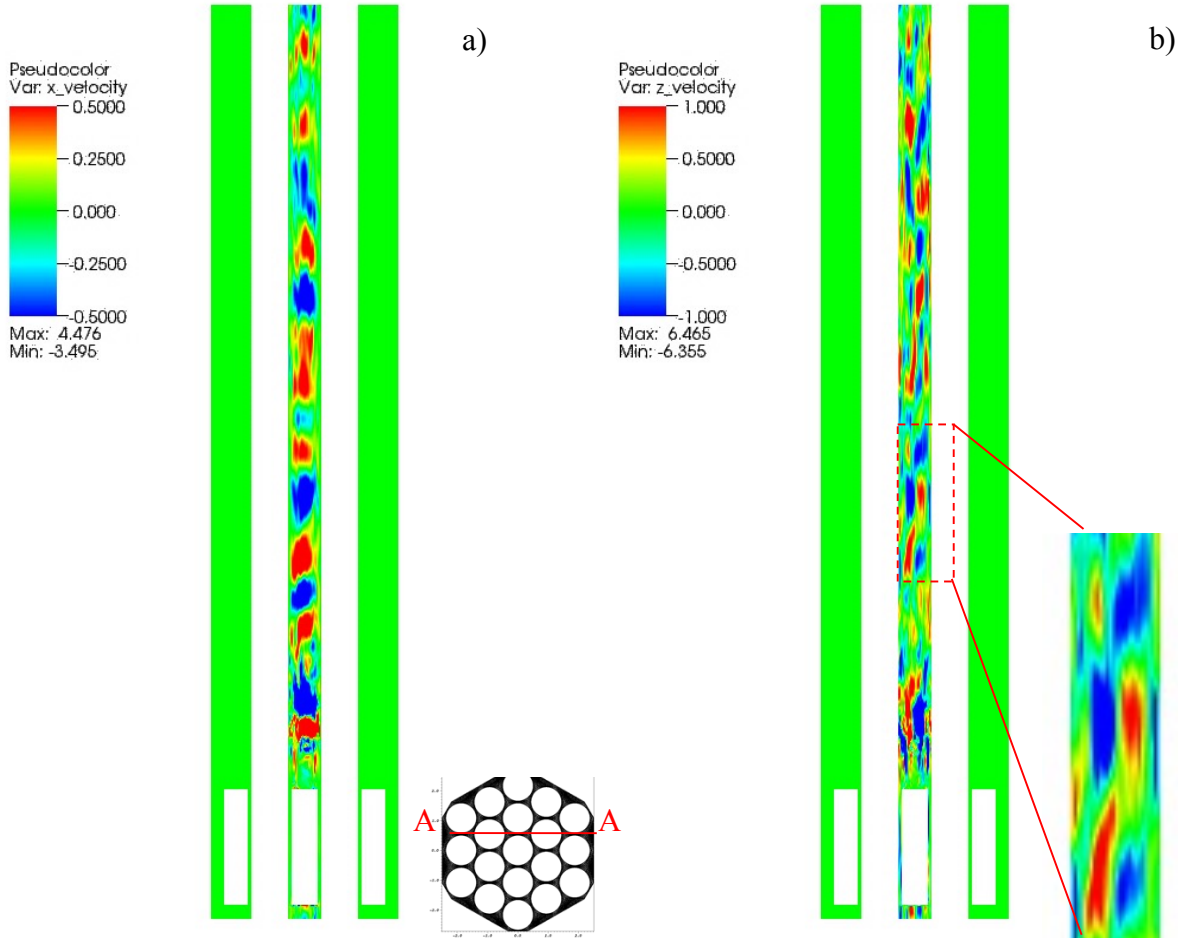

Figure 10. 19-pin SFR rod bundle, principal POD mode: (a) cross velocity; (b) streamwise velocity with detail.

## 3.2. Jet Mixing and Thermal Striping

In this section we describe two experiements, with particular attention given to thermal striping.

### 3.2.1 MAX Experiment

Electric power plants rely on piping systems to transport and sometimes mix hot and cold fluid streams. Mixing may generate local fluctuations in fluid temperature that can be transferred to the piping itself to produce thermal cycling and thermal stresses. Cycling may be random or periodic, but in either case it can progressively weaken materials through the process of thermal fatigue. High-cycle thermal fatigue has been observed in boiler tubes, a steam exhaust silencer, and mixing tees of nuclear power stations. Even modest thermal cycling can critically weaken materials over the decades-long life of a power plant.

The term "thermal striping" is often used for the collective phenomena of fluctuating fluid temperature, thermal cycling of structures, and associated stresses. Some researchers also include the mechanical fatigue induced by protracted cycling. Thermal striping can occur in regions of thermal stratification, but here our focus in on convective mixing as the source of thermal cycling.

Thermal striping is of special concern for liquid metal-cooled nuclear fast reactors. Fluctuations in fluid temperature are readily transferred to piping components because of the low Prandtl number of liquid metals. Fluctuations can be substantial because of potentially high $\Delta T$ between mixing flow streams, in some instances >100 $^{o}$C. For example, weld failures in sodium-cooled fast reactors (SFRs) have been associated with thermal striping. A general sense of the many potential sources of thermal striping in fast reactors is available in [12].

Various thermal mixing experiments have been conducted in the context of thermal striping. Of particular interest are those employing liquid sodium since it is the working fluid of the most established class of fast reactor. Hot and cold concentric sodium jets have been mixed in tanks with measurements of the temperature field for comparison with air and water systems in order to characterize similarity relations [13]. Kimura et. al. [14] examined heat transfer between a wall and three rectangular jets with the central one colder than the outer two. Temperature fluctuations in both the fluid and wall were measured in order to characterize transient heat transfer.

Sodium experiments provide considerable insight into thermal striping mechanisms but are of limited utility for CFD validation. Such validation should include a direct comparison with experimentally measured velocity fields, but the opacity of sodium precludes high-fidelity velocity mapping with standard optical measurement techniques such as PIV. Likewise, optical temperature-mapping techniques such as laser-induced fluorescence are unsuitable for sodium. As a result, transparent fluids such as air or water are used in lieu of sodium when the data is intended for rigorous CFD benchmarking. We examine here work performed as part of the MAX experiment [15], which illustrates a successful validation program in this area. MAX is a jet mixing experiment with wall impingement in a geometry patterned off the upper plenum of a sodium fast reactor. It mimics situations where temperature mismatches in core assembly outlet flows impinge on structures and induce thermal cycling. The working fluid is air rather than sodium, in order to accommodate optical instrumentation.

In MAX a small change in the inlet geometry for the hexahedral jets substantially altered both the flow and temperature fields. The impingement flow field varied with inlet configuration: three stagnant regions directly above the inlets for the extended case (e.g., jets extended in the domain) but only one for the flush case (e.g., jets flushing at the domain boundary). In the extended case, flow field root mean square (RMS) is relatively high and concentrated in a narrow band between jets, whereas it is more diffuse for the flush case. This collection of data provided an indirect indication of jet stability in the extended configuration and instability for the flush case.

A similar dichotomy was seen in measurements of air temperature across the impingement flow field during non-isothermal tests. Measurements were made at Re=10000 and inlet $\Delta T$ of 4 $^{o}$C. For the flush case, temperature gradients across the impingement region are relatively small. This result is consistent with enhanced mixing driven by instabilities that wash out the gradients before they reach the lid. The extended case exhibits more sharply defined hot and cold spots along with a narrow band of relatively high RMS similar to the band seen for the RMS velocity. This is consistent with stable jets and inhibited mixing so that much of the inlet $\Delta T$ is maintained all the way to the lid.

Nek5000 was used to simulate the flow field [15] in both the flush and extended cases under isothermal conditions. It reproduced the lid flow patterns seen in the PIV data (Fig. 11 and Fig. 12), including the number of stagnant spots and their positions over the inlets. In quantitative comparisons, the velocity magnitude matched well in the extended case but was less successful in the unstable flush case, a result that was not unexpected since the experiment itself exhibited appreciable variability in the flush case. Simulations corroborated experimental data, indicating the flow dichotomy is not the result of a chance combination of jet core length and distance to the lid but, rather, the difference in inlet geometry. The clear flow dichotomy exhibited by this two-jet setup presents a simple case for testing CFD tools and their ability to predict flow field changes driven by subtle changes in geometry in the context of thermal striping.

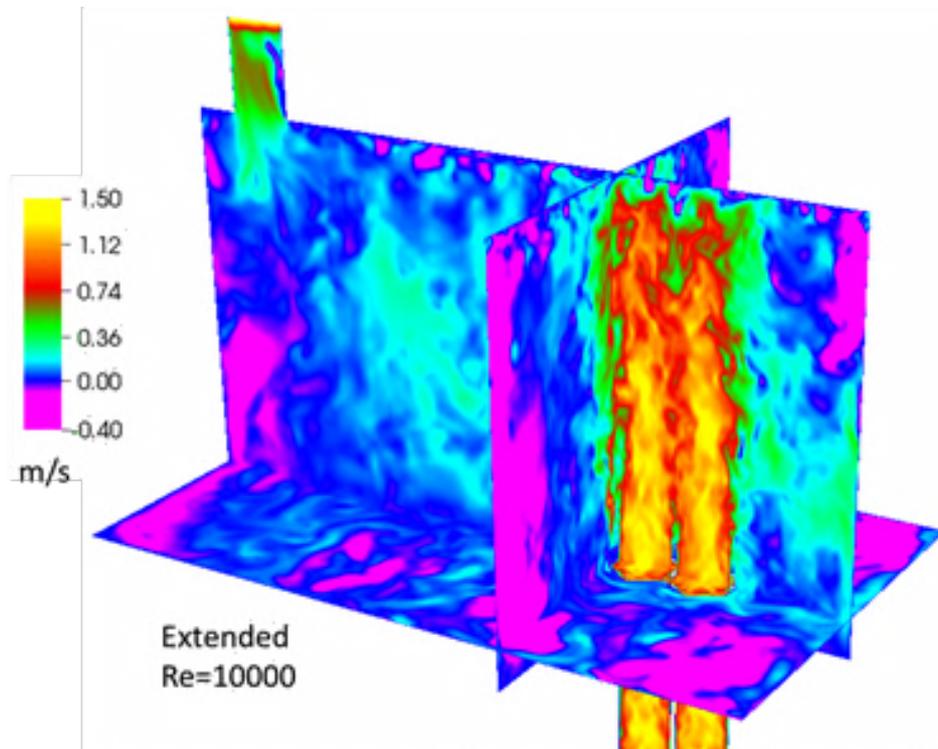

Fig. 11. Instantaneous vertical velocity (Nek5000) across midplane, 100 mm above base, and x=-136 (west edge of west inlet).

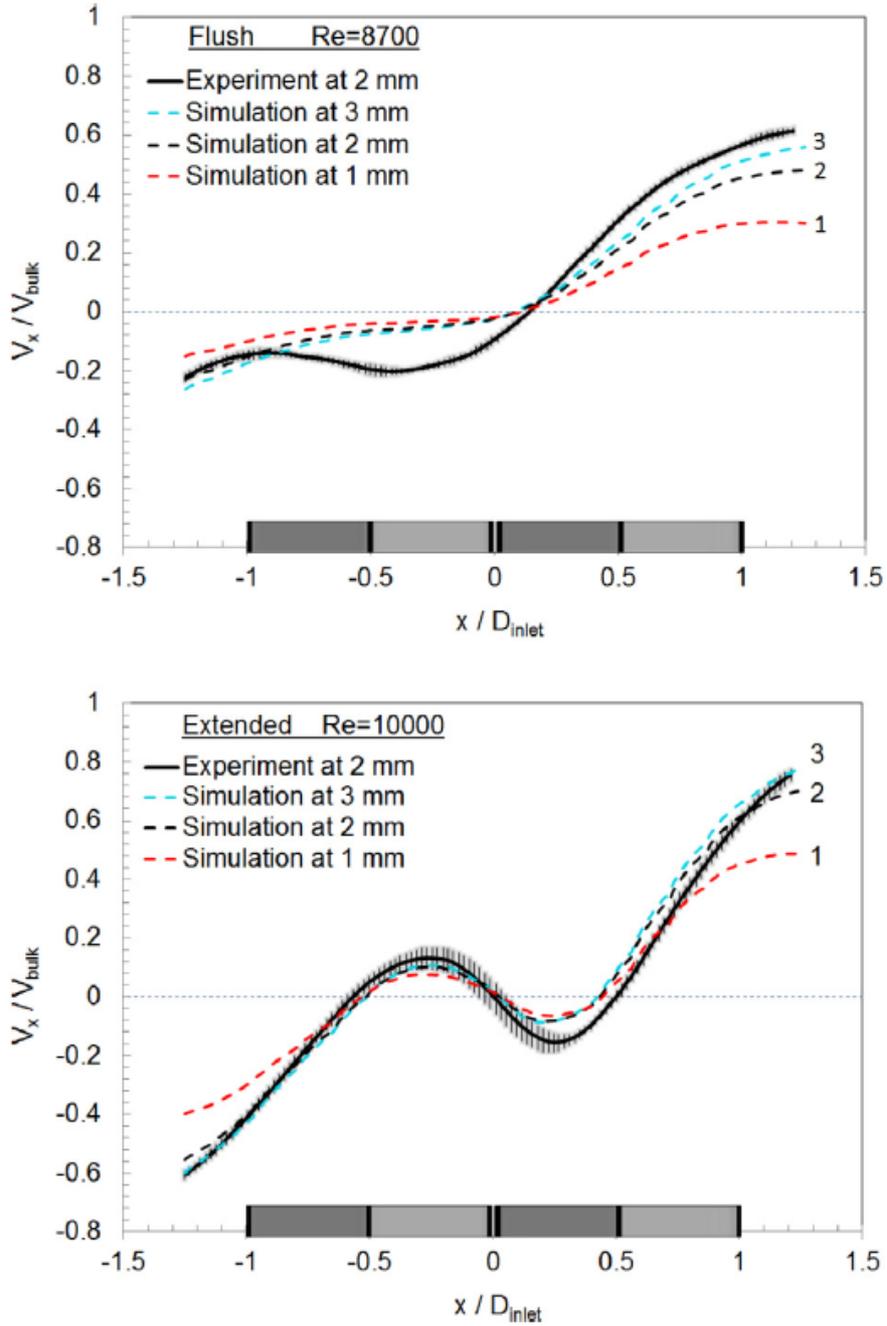

Fig. 12. Comparison between simulation and experiment: velocity below lid along tank centerline for isothermal flow. Top: flush case at Re=8700. Bottom: Extended case at Re=10000. Experimental data at 2 mm from wall. Simulation data at three wall distances.

3.2.2 PLAJEST Experiment

As mentioned in the preceding section, several attempts have been made at measuring and qualifying CFD codes for thermal striping in sodium. Kimura et al. [14] performed a sodium experiment, called PLAJEST, that had triple-parallel jets along a stainless steel wall. The authors showed that the transfer characteristics of temperature fluctuation from fluid to structure could be evaluated by using a heat transfer coefficient obtained from a transfer function between temperature fluctuations in the fluid and structure. Figure 13 presents a volume rendering plot of the instantaneous temperature as simulated in Nek5000 for the PLAJEST experiment.

Tokuhiro et al. [16] performed a water experiment, called WAJECO, of the triple-parallel jets mixing phenomena and evaluated the mixing process among the jets. The attenuation process of the temperature fluctuation from fluid to structure was important for predicting the thermal striping phenomena. Furthermore, in the water experiment WAJECO with the same configured test section as in the sodium experiment PLAJEST, a stainless steel plate with thermocouples was set along the flow in order to evaluate the characteristics of temperature fluctuation transfer from fluid to structure in the two experiments.

An extensive boundary condition and mesh sensitivity study with the WAJECO/PLAJEST simulation demonstrated that LES methods can be used to simulate the flow in this geometry [17], whereas URANS and RANS methods tend to underestimate the mixing and lead to poor results for both velocity and temperature predictions, especially downstream.

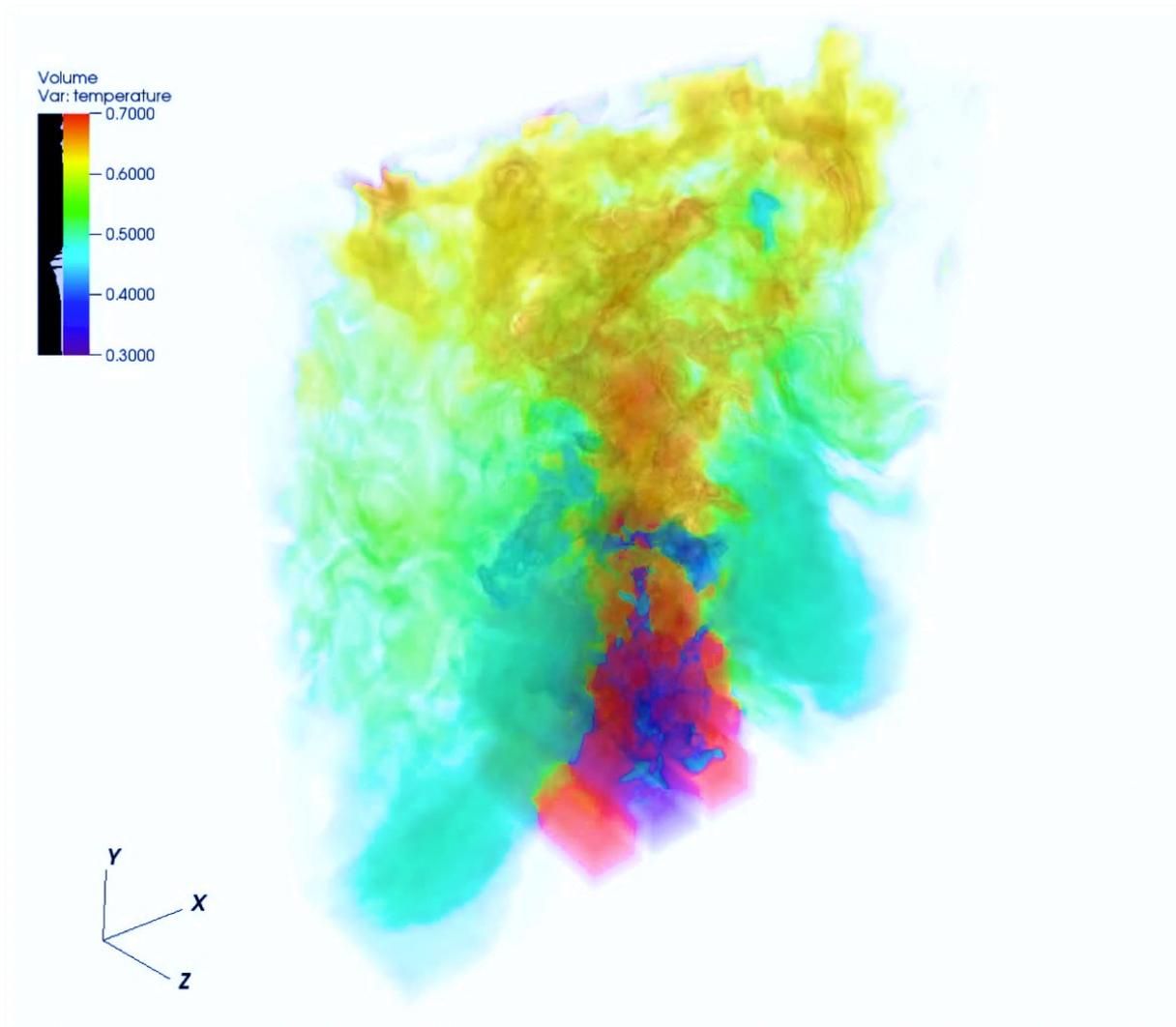

Figure 13. Volume rendering of the instantaneous temperature distribution in PLAJEST, showing Nek5000 results.

**3.3. Full-Core Applications**

In order to design an inherently safe fast reactor, reactivity dependence on radial core expansion must be engineered into the reactor plant to ensure a loss of reactivity during transient events. In advanced SFR concepts under consideration in the United States, the core is designed to bow outward in response to thermal expansion of the structures in any transient where the core is heating. The grid plate and load pads, which support the core from below and restrain it from the top, respectively, also expand outward. Moreover, the core restraint system is designed such that the fuel assemblies bow outward in the middle, further separating the fuel pins. When controlled correctly, core expansion causes the fuel assemblies to move farther apart from each other, which has a negative reactivity effect and helps shut down the reactor. Simulation of this expansion, which is essential to the safety of these reactor concepts, necessitates the coupling of structural mechanics, thermal hydraulics, and structural mechanics. The primary thermal-hydraulic analysis needed to simulate the full-core response is through the temperature in the ducts. Reduced-order models such as porous media or subchannel analysis are good candidates for simulating the full-core behavior.

Porous media modeling for wire-wrapped rod bundles has long been applied, but it has not been traditionally applied to temperature predictions inside the assembly ducts with explicit modeling of the ducts. In order to evaluate porous media models for ducted fuel assemblies with wire-wrappers, reference simulations were performed with a CFD model. The geometry was explicitly represented in the CFD model, with a 61-pin wire-wrapped fuel bundle; and three porous media models were evaluated. Results are summarized in Figure 14. The 3 region model best accounts for the variation in the ratio of solid volume to coolant volume in different parts of the assembly [18]. The results of the 2 region porous media model reproduce the duct temperature of CFD simulation with a good precision. However, the uniform model, which is frequently used in other analyses, exhibits poor agreement.

The advanced nuclear reactor modeling and simulation toolkit SHARP was employed with the CFD model to perform a first-of-a-kind analysis of the core radial expansion phenomenon in a full-core SFR [19]. Physics models of a full-core model have also been developed. A fully integrated quasi-static simulation of a full-core test problem has also been performed. An image of the full-core deformation is shown in Figure 15.

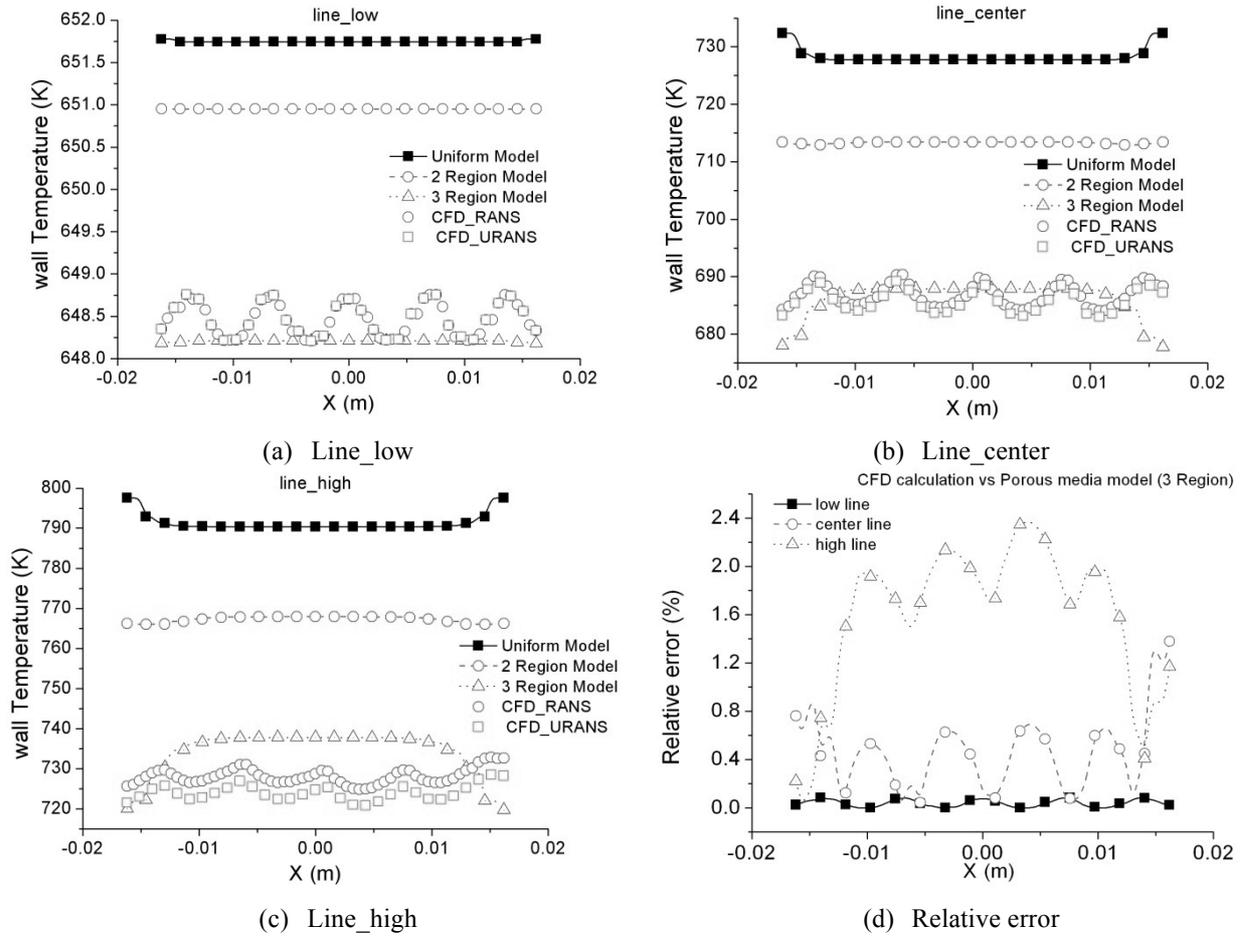

Figure 14. Comparisons of duct temperature between the CFD model and porous media models. Plots are at different heights (a–c) and relative error. The accuracy can be further improved by adding additional terms to the porous media model [18].

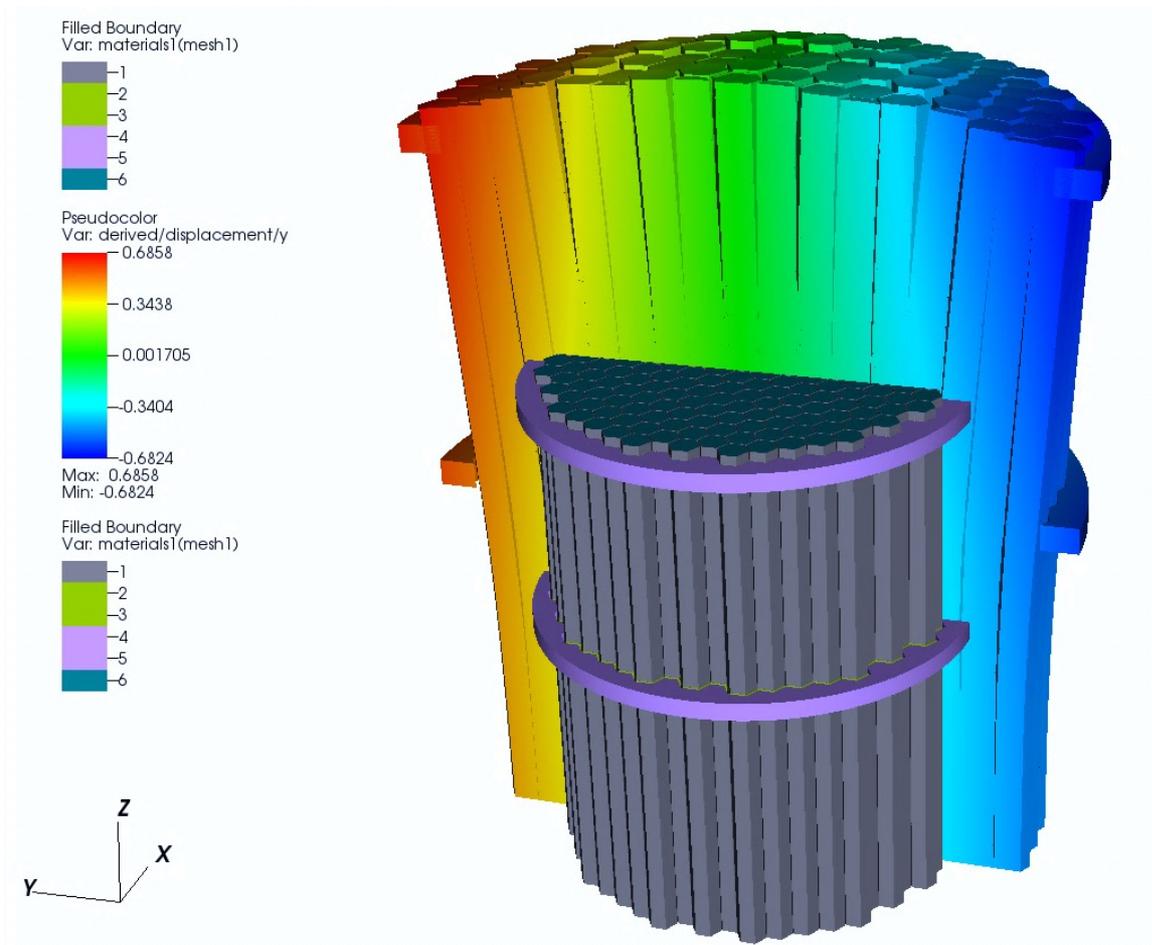

Figure 15. Magnified (100x) displacements colored by the displacement in the y direction for a full-core simulation of an SFR.

## 4. Conclusions

As part of the NEAMS program, the Nek5000 spectral element code was applied to liquid metal reactor flows. Here we have examined some of the foundations of Nek5000 and its application to liquid metal flows. Particular focus has been placed on rod bundle flows, but we have also dedicated significant attention to thermal striping applications. We have also demonstrated the use of Nek5000 for coarse CFD applications with a full-core porous media coupled simulation to predict radial core expansion.

## Acknowledgments

This material was based upon work supported by the U.S. Department of Energy, Office of Science, under contract DE-AC02-06CH11357.